\documentclass[aps,pra,twocolumn,groupedaddress,showpacs]{revtex4-1}
\usepackage{color,graphicx}
\usepackage{dcolumn}
\usepackage{bm}

\begin{document}

\title{Sub-Doppler cooling of sodium atoms in gray molasses}
\author{Giacomo Colzi$^{1,2}$}
\author{Gianmaria Durastante$^{1}$}
\altaffiliation{present address: Institut f\"ur Quantenoptik und Quanteninformation, \"Osterreichische Akademie der Wissenschaften, 6020 Innsbruck, Austria}
\author{Eleonora Fava$^{1}$}
\author{Simone Serafini$^{1}$}
\author{Giacomo Lamporesi$^{1,2}$}
\author{Gabriele Ferrari$^{1,2}$}

\affiliation{$^1$ INO-CNR BEC Center and Dipartimento di Fisica, Universit\`a di Trento, 38123 Povo, Italy}
\affiliation{$^2$ Trento Institute for Fundamental Physics and Applications, INFN, 38123 Povo, Italy}

\date{\today}

\begin{abstract}
We report on the realization of sub-Doppler laser cooling of sodium atoms in gray molasses using the D1 optical transition ($3s\, ^2S_{1/2} \rightarrow 3p\, ^2P_{1/2}$) at 589.8 nm. The technique is applied to samples containing $3\times10^9$ atoms, previously cooled to 350 $\mu$K in a magneto-optical trap, and it leads to temperatures as low as 9 $\mu$K and phase-space densities in the range of $10^{-4}$. The capture efficiency of the gray molasses is larger than 2/3, and we observe no density-dependent heating for densities up to $10^{11}$ cm$^{-3}$.
\end{abstract}

\maketitle
\section {Introduction}
Sub-Doppler laser cooling has been a well-established and widespread technique since the late 1980s \cite{lett88}. It consists in using light polarization gradients and optical pumping to cool down atoms below the Doppler temperature limit $T_{\mathrm{D}}=\hbar \Gamma / 2 k_{\mathrm{B}}$  \cite{Dalibard89,Ungar89}, where $\hbar$ is the reduced Planck constant, $k_{\mathrm{B}}$ is the Boltzmann constant and $\Gamma$ is the natural linewidth. This technique, following a Doppler precooling stage in a standard magneto-optical trap (MOT), greatly improves the efficiency of the cooling process within its velocity capture range, but fundamentally does not remove the heating effects associated with the stochastic nature of photon absorption and spontaneous emission cycles. In addition, the atomic sample density remains limited both by light-assisted collisions \cite{Prentiss88,Marcassa93} and by effective atom-atom repulsive forces caused by reabsorption of scattered photons \cite{Walker90}. Improvements are obtained in schemes where slowed atoms are preserved from the above mentioned effects through selective trapping into dark states---states that are not coupled with the light field. Examples include the dark-spot (DS) -MOT scheme \cite{Ketterle93} and schemes where the fraction of atoms optically pumped in a dark hyperfine state is dynamically controlled \cite{Landini11}. The use of dark states in the achievement of lower and lower temperatures clearly leads to a gain in evaporation efficiency, hence in the final atom number. The advantages offered by dark states can also be exploited in other contexts where spontaneous emission still plays a pivotal role, as exemplified by the recent developments in single-site resolved quantum-gas microscopy \cite{Cheuk15,Haller15}. For this reason, studying techniques that take advantage of atomic dark states is still a promising field of study. This holds especially true given the fact that highly degenerate quantum gases shifted rapidly from being the object of study to being an experimental tool in addressing fundamental problems in physics and in devising cold-atom based quantum simulators \cite{Bloch12}.

Velocity-selective coherent population trapping (VSCPT) \cite{Aspect88} allows one to cool atoms even below the photon recoil temperature limit $T_{\mathrm{rec}}= \hbar^2 k_{\mathrm{L}}^2 / m k_{\mathrm{B}} $ (where $k_{\mathrm{L}}$ is the wave number and $m$ is the atomic mass), taking advantage of electromagnetically induced transparency \cite{Alzetta76,Arimondo76}. Its main limitation is
that it relies on the diffusion in momentum space, due to spontaneous emission, accumulating a relatively small population in a velocity-selective dark state. Gray molasses (GM) cooling \cite{ Shariar93,Grynberg94,Weidemuller94}, on the other hand, consists of polarization gradient cooling in the presence of VSCPT: dark states are given by a coherent superposition of degenerate Zeeman sublevels in the ground hyperfine manifold, while states coupling with the radiation field are shifted in energy according to polarization gradient effects. In such a scheme atoms are continuously optically pumped into the dark states as they cede kinetic energy to the light field. The velocity-selective mechanism is provided by non adiabatic transitions: faster atoms in dark states have a higher transition probability to bright states from which the cooling cycle reiterates. Dark states exist only for $\left| F \right\rangle \rightarrow \left|F-1 \right\rangle$ or $\left| F \right\rangle \rightarrow \left| F \right\rangle$ transition; the  cooling mechanism works for blue detunings only, meaning that atoms in velocity classes outside the sub-Doppler capture range are heated up by Doppler heating. In addition, stray magnetic fields have to be compensated to preserve the cooling efficiency.
In the presence of magnetic fields and for proper light field configurations also a different cooling and trapping scheme can be devised. This is referred to as dark optical lattice (DOL) \cite{Grynberg94,Pestas96}. In the strong magnetic field regime, for which the Zeeman energy shift is higher than the polarization gradient \cite{Hemmerich95}, it consists in lifting the Zeeman degeneracy such that dark states correspond to $m_F$ eigenstates uncoupled to the light field due to its local polarization.

\begin{figure*}[t!]
	\centering
	\resizebox{0.95\textwidth}{!}{
		\includegraphics{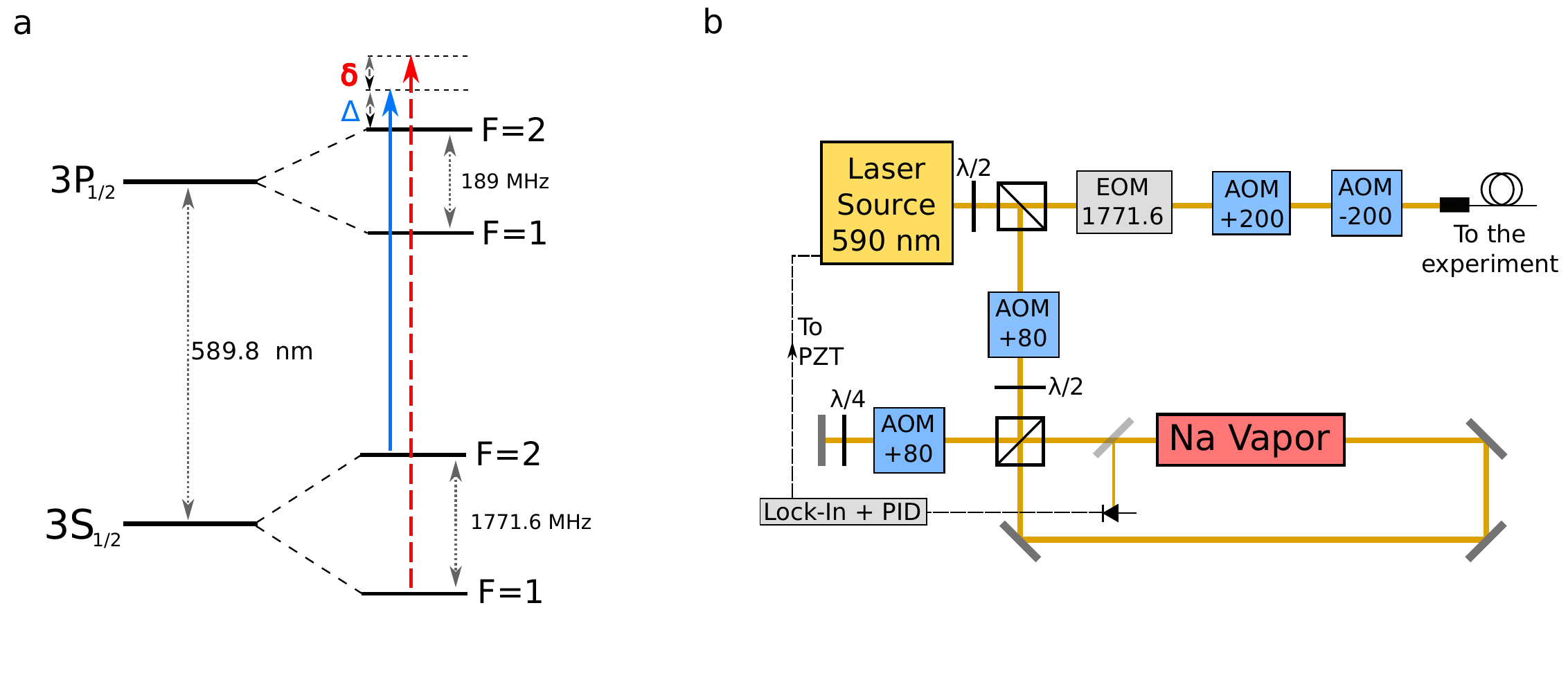}}
	\caption{\label{Figure1} \textbf{(a)} Sodium $\mathrm{D_1}$ level structure. Cooler (blue, solid) and repumper (red, dashed) transitions are shown. \textbf{(b)} Sketch of the $\mathrm{D_1}$ laser source.}
\end{figure*}

GM cooling was achieved experimentally in the 1990s on $^{133} \mathrm{Cs}$ \cite{Valentin92,Boiron95,Boiron96,Triché99} and on $^{87} \mathrm{Rb}$ \cite{Esslinger96} atoms using the $\mathrm{D_2}$ transition $(\mathrm{nS}_{1/2} \rightarrow \mathrm{nP}_{3/2})$. The transverse cooling of a $^{52} \mathrm{Cr}$ atomic beam \cite{Drewsen96} was also reported. The $\mathrm{D_1}$ transition $(\mathrm{nS}_{1/2} \rightarrow \mathrm{nP}_{1/2})$ was already used in the context of DOLs with $^{87} \mathrm{Rb}$ \cite{Hemmerich95} and $^{85} \mathrm{Rb}$ \cite{Stecher97}. This choice is beneficial due to the absence of off-resonant excitations on the blue side of the hyperfine manifold of interest and to the more resolved energy spectrum. More recently, GM cooling was demonstrated on  $^{85} \mathrm{Rb}$ \cite{Elman05}, $^{40}\mathrm{K}$ \cite{Fernandes12, Sievers15}, $^{7}\mathrm{Li}$ \cite{Grier13}, $^{39}\mathrm{K}$ \cite{Salomon13,Nath13}, and $^{6}\mathrm{Li}$ \cite{Burchianti14, Sievers15} using the $\mathrm{D_1}$ transition, and on metastable $^{4}\mathrm{He}$ \cite{Bouton15}.  

In this article we report on an experimental characterization of GM cooling of $^{23} \mathrm{Na}$ atoms exploiting the $\mathrm{D_1}$ transition $\left| F_g=2 \right\rangle \rightarrow \left| F_e=2 \right\rangle$. The level structure is shown in Fig. \ref{Figure1}(a). 
In the following section (Sec. \ref{expapp}) the reader will find the description of the $\mathrm{D_1}$ laser source. Subsequently (Sec. \ref{char}) the description of the experimental procedure and its characterization will be provided, before drawing the conclusions in the last section (Sec. \ref{conc}).

\section {\label{expapp}Experimental apparatus}
The $\mathrm{D_1}$ laser source employed for GM cooling is sketched in Fig. \ref{Figure1}(b). The main $590 \; \mathrm{nm}$ laser source comprises a $1180 \, \mathrm{nm}$ master oscillator diode (Innolume GC-1178-TO-200) mounted in Littrow configuration \cite{Ricci95} and capable of emitting $50 \; \mathrm{mW}$, a Raman fiber amplifier (RFA) pumped with an ytterbium fiber laser capable of emitting up to $8 \; \mathrm{W}$ of power while maintaining the spectral properties of the master oscillator beam, and a resonant second harmonic generation stage based on a $15 \; \mathrm{mm}$ $\mathrm{Li B_3 O_5}$ nonlinear crystal mounted in a bow-tie cavity stabilized by means of polarization spectroscopy \cite{Hansch80}.

Operating the RFA at $5 \; \mathrm{W}$ we obtain $2.2 \; \mathrm{W}$ of power at $590 \; \mathrm{nm}$. The laser frequency is tuned by acting on the master laser, and is stabilized on the  $\left| F_g=2 \right\rangle \rightarrow \left| F_e=2 \right\rangle$ cooler transition frequency using modulation-transfer saturated absorption spectroscopy on a heatpipe containing sodium vapors. The repumper laser frequency, corresponding to $\left| F_g=1 \right\rangle \rightarrow \left| F_e=2 \right\rangle$  is obtained generating frequency sidebands in the laser field by means of an electro-optic modulator (EOM) driven at about $1771.6 \; \mathrm{MHz}$, the hyperfine splitting frequency of the electronic ground state. After two acousto-optic modulators (AOMs)  acting as fast switches, a polarization maintaining optical fiber is used to deliver the laser light on the optical table hosting the atomic cell.

The aforementioned experimental configuration enables us to control all the parameters relevant to GM cooling. Both the cooler detuning $\Delta$ and the total intensity $I_\mathrm{C}$ + $I_{\mathrm{rep}}$ (cooler and repumper intensity, respectively) delivered to the atoms are tuned by changing the frequency and amplitude of the RF signals driving the two AOMs before the optical fiber. The relative intensity  $I_{\mathrm{rep}}/{I_\mathrm{C}}$ and the Raman detuning $\delta$ between repumper and cooling can be tuned by changing the amplitude and frequency of the microwave field driving the EOM. 
\begin{figure*}[t!]
	\resizebox{\textwidth}{!}{
		\includegraphics{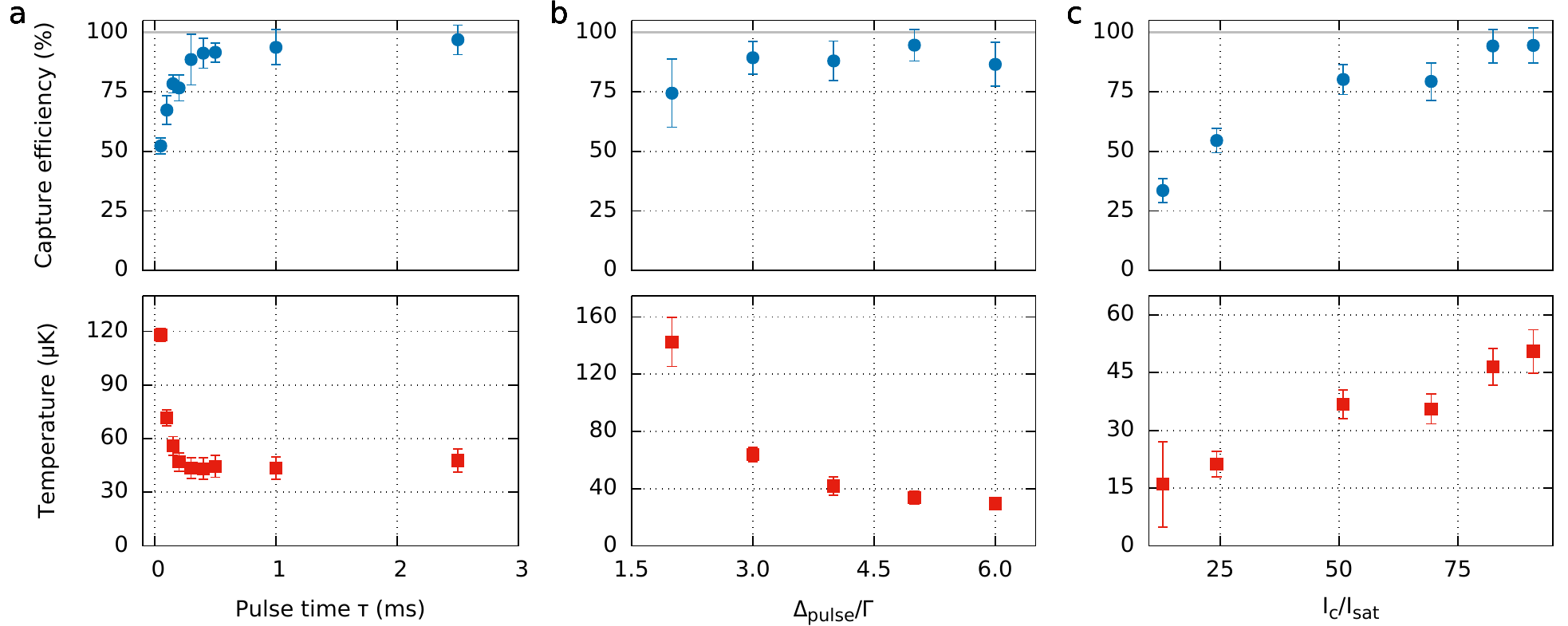}}
	\caption{\label{Pulsechar}(a) Capture efficiency (blue circles) and temperature (red squares) after a single GM pulse at constant intensity $I_C=80 \; I_{\mathrm{sat}}$ and fixed detuning $\Delta_{\mathrm{pulse}}=4\Gamma$, as a function of the pulse duration $\tau$, (b) as a function of the detuning $\Delta_{\mathrm{pulse}}$ with $\tau=0.5 \; \mathrm{ms}$ and $I_C=80 \; I_{\mathrm{sat}}$, and (c) as a function of the total cooler intensity $I_C$  with $\tau=0.5 \; \mathrm{ms}$ and $\Delta_{\mathrm{pulse}}=5\Gamma$. Error bars correspond to one standard deviation. The repumper relative intensity and detuning are fixed to $I_{\mathrm{rep}}/I_{\mathrm{C}}\simeq 4 \% $, and $\delta = 0$, respectively.}
\end{figure*}
We deliver up to $200 \; \mathrm{mW}$ of power to the atoms starting from a single collimated beam of $4.7 \; \mathrm{mm}$ waist ($e^{-2}$ radius). The corresponding intensity is about $90 \, I_{\mathrm{sat}}$ in units of the $\mathrm{D_2}$ saturation intensity for $\sigma^{\pm}$ polarized light $I_{\mathrm{sat}}=6.26 \; \mathrm{mW / cm^2}$. This beam is superimposed on the one generated by the $\mathrm{D_2}$ laser source that is used for standard MOT and bright molasses (BM) cooling (for a detailed description, see Ref. \cite{Lamporesi13}). The two copropagating beams are then split into six independent beams and sent on the atoms along three orthogonal directions, counterpropagating in pairs with $\sigma^+ - \sigma^-$ polarizations. 
The atoms can be loaded in a DS-MOT using an additional $\mathrm{D_2}$ repumper beam that is sent through a circular obstacle. Imaging the obstacle on the atoms results in a hollow repumper beam with a sharp hole of radius $4 \; \mathrm{mm}$.

\section{\label{char}Experimental procedure and system characterization}

\begin{figure} [h]
	\resizebox{\columnwidth}{!}{
		\includegraphics{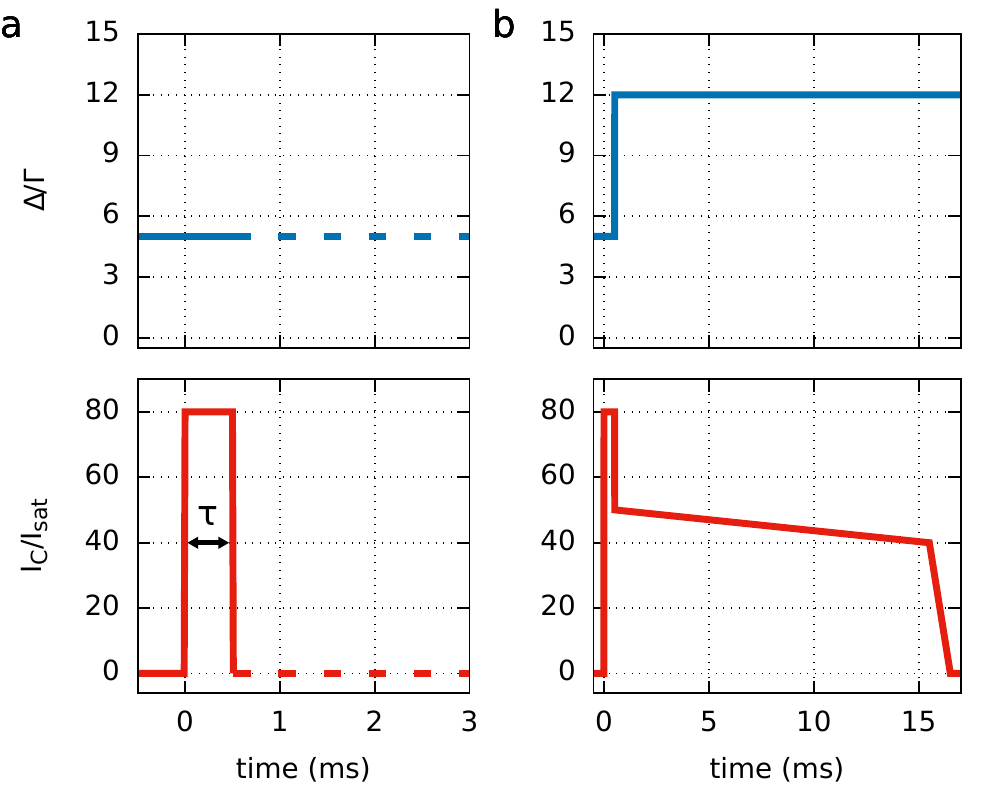}}
	\caption{\label{seq}(a) Cooler detuning $\Delta/\Gamma$ (blue line) and intensity $I_{\mathrm{C}}/I_\mathrm{sat}$ (red line) as a function of time for the single GM pulse, and (b) for the GM sequence with ramp. Shown intensity values already take into account the reduction of AOMs diffraction efficiency for high detuning.}
\end{figure}

Sodium atoms are loaded in the DS-MOT before applying GM cooling. The loading sequence lasts about $8 \; \mathrm{s}$, yielding $ \sim 3 \, \mathsf{x} \, 10^{9}$ atoms at a temperature of $T \sim 350 \; \mathrm{\mu K}$. GM cooling sequence is initiated $200 \; \mathrm{\mu s}$ after DS-MOT laser beams and magnetic fields are turned off. This ensures the decay of transient magnetic fields to below $20 \; \mathrm{mG}$.
At the end of GM sequence we perform absorption imaging in time of flight (TOF), from which we measure the number of atoms and their temperature in the plane orthogonal to the imaging axis. The central phase-space density of the sample is calculated as \textit{PSD}$= n_{\mathrm{p}}\lambda_{\mathrm{T}}^{3}$, where $n_{\mathrm{p}}$ is the peak density of the Gaussian distribution observed \textit{in situ} and $\lambda_{\mathrm{T}}$ is the thermal de Broglie wavelength. The capture efficiency of GM is calculated normalizing the atom number after GM to the average atom number right after the DS-MOT sequence.

\subsection{Constant Intensity Pulse}
We first study the effect of a GM stage made with a constant intensity pulse of duration $\tau$ with laser beams kept at fixed detuning $\Delta_\mathrm{pulse}$. 

The capture efficiency and the temperature after the pulse are reported as a function of $\tau$ in Fig. \ref{Pulsechar}(a). Nearly the whole sample is cooled down to a temperature of $T\simeq 45 \; \mathrm{\mu K}$ in about $0.5 \; \mathrm{ms}$. The other parameters are fixed $\Delta_{\mathrm{pulse}}=4 \Gamma$ ($\Gamma= 2\pi \, \mathsf{x}  \, 10 \;\mathrm{MHz}$) and $I_\mathrm{C}= 80 \, I_{\mathrm{sat}}$. The repumper sideband intensity is set to a few percent ($ \simeq 4 \% $) of the cooler intensity and $\delta$ is set to zero.
 
\begin{figure*}[t!]
	\resizebox{\textwidth}{!}{\includegraphics{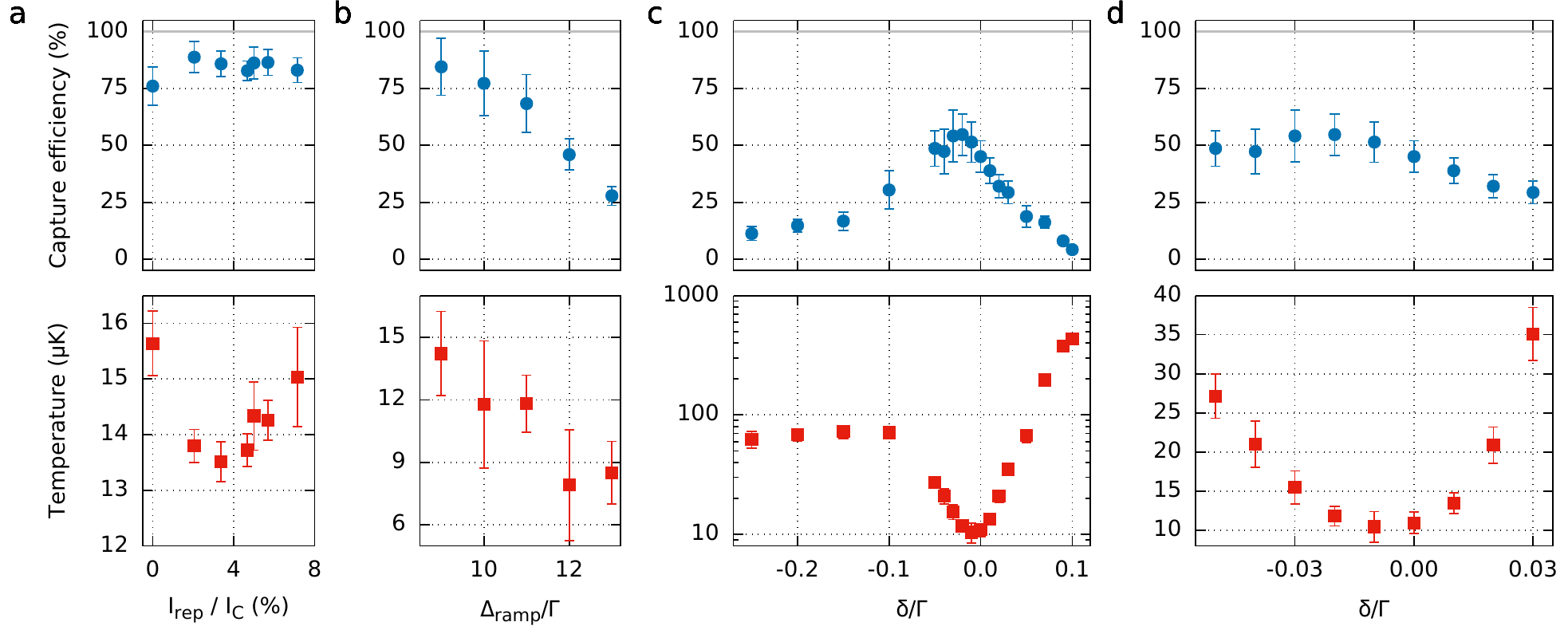}}
	\caption{\label{raman}(a) Capture efficiency (blue circles) and temperature (red squares) after the complete ramp sequence characterized as a function of $I_\mathrm{rep} / I_{\mathrm{C}}$ for $\delta=0$ and $\Delta_{\mathrm{ramp}}=10\Gamma$, (b) as a function of the ramp detuning $\Delta_{\mathrm{ramp}}$ for $\delta=-0.01 \Gamma$ and $I_\mathrm{rep} / I_{\mathrm{C}}=0.034$, and (c), (d) as a function of $\delta$ for $\Delta_{\mathrm{ramp}}=12\Gamma$ and $I_{\mathrm{rep}}/ {I_\mathrm{C}}= 0.034$. Error bars correspond to one standard deviation. The initial pulse parameters are fixed $\tau=0.5 \; \mathrm{ms}$, $\Delta_{\mathrm{pulse}}=5\Gamma$, and $I_\mathrm{C}= 80 \; I_\mathrm{sat}$} 
\end{figure*}

Figure \ref{Pulsechar}(b) shows that the capture efficiency is maximized for $\Delta_{\mathrm{pulse}} \simeq 5\Gamma$ (in case of $\tau=0.5 \; \mathrm{ms}$).

In Fig. \ref{Pulsechar}(c) the dependence on the cooler intensity is shown. Both the temperature and the capture efficiency decrease linearly with the pulse intensity. This behavior is expected in the general framework of polarization-gradient cooling in which the temperature lower bound scales as $I/\Delta_{\mathrm{pulse}}$ and the capture range decreases accordingly with intensity. Atoms outside the capture range are thus excluded from the sub-Doppler cooling process and hence lost. Because of this we opted for the  sequence described in Sec.\ref{Ramp}.

\subsection{\label{Ramp}Intensity Ramp}
The entire sequence was optimized piecewise, as sketched in Fig. \ref{seq}. A pulse of $\tau=0.5 \; \mathrm{ms}$, an intensity value of $I_\mathrm{C}= 80 \, I_{\mathrm{sat}}$, and detuning $\Delta_{\mathrm{pulse}}=5\Gamma$ is applied to maximize the capture efficiency. Then the cooler detuning is increased and the intensity is reduced to keep most of the atoms in the velocity capture range of the sub-Doppler process while the temperature decreases.  $I_{\mathrm{rep}}/I_{\mathrm{C}}$ is set to $3.4 \%$; see Fig. \ref{raman}(a). The cooler detuning is increased to $\Delta_{\mathrm{ramp}}=12 \Gamma$. This choice maximizes the final \textit{PSD} even if it reduces the capture efficiency of the process, as shown in Fig. \ref{raman}(b). Due to the reduction of diffraction efficiency in AOMs, the total intensity reduces to about $I_\mathrm{C}= 50 \, I_{\mathrm{sat}}$. At this stage we observe that a linear ramp of the detuning, instead of a sudden change, does not produce significant improvements. The intensity is linearly reduced to $40 \, I_{\mathrm{sat}}$ in $15 \; \mathrm{ms}$ and finally to zero in $1 \; \mathrm{ms}$. Reducing the duration of the first ramp to 8 ms does not alter the overall performances. The complete sequence is plotted in Fig. \ref{seq}(b).
Figure \ref{raman}(c) shows that in the vicinity of the Raman resonance the capture efficiency is maximized, while for positive $\delta$ above $2\pi \, \mathsf{x} \, 200 \; \mathrm{kHz} $ a large heating sets in. 
The best performance of the process both in temperature and capture efficiency is obtained for small negative Raman detuning $\delta \simeq - 2\pi \, \mathsf{x} \; 100 \; \mathrm{kHz}$, as shown in Fig. \ref{raman}(d).
As discussed in Refs. \cite{Grier13,Nath13,Sievers15}, close to the Raman resonance a long-lived dark manifold involving both hyperfine ground levels emerges, further suppressing the scattering rate thus improving the efficiency of GM cooling. Moreover, additional polarization gradient cooling (or heating) cycles take place between dressed states of both manifolds. Whether cooling or heating processes are favored depends on $\delta$ and on the ratio $I_{\mathrm{rep}}/I_{\mathrm{C}}$. Slightly out of resonance a subnatural width Fano profile in the transition probability can manifest due to the interference of different excitation processes \cite{Lounis92}, leading to strong heating of the sample. In the regime $I_{\mathrm{rep}}/I_{\mathrm{C}} \ll 1 $, consistent with our results, a heating peak is expected for $\delta>0$. 

\subsection{Final Results}
\begin{table}[h]
	\begin{center}
		\caption{\label{resfin} Comparison between the different cooling stages results and the initial condition after DS-MOT. Eff$\%$: approximative capture efficiency; $n_{\mathrm{p}}$: peak density; $T$: temperature; \textit{PSD}: phase-space density. The parameters chosen for GM operation are the same as in Fig. \ref{seq}, with $\delta= -0.01 \Gamma$ and $I_\mathrm{rep}/I_{\mathrm{C}}=3.4 \%$.} 
		\begin{ruledtabular}
		\begin{tabular}{lllll} 
			$\space$ & Eff$\%$ & $n_{\mathrm{p}} \, ( \mathrm{cm^{-3}})$ & $T \, (\mu \mathrm{K})$ & \textit{PSD} \\
			\hline\noalign{\smallskip}
			DS-MOT & -- & $9(1) \, \mathsf{x} \ 10^{10} $ & $355(12)$ & $6.8(10) \, \mathsf{x} \, 10^{-7}$ \\
			$D_2$ BM & $80$ &  $4.3(2)\, \mathsf{x} \, 10^{10} $ & 38.1(9) &  $8.9(6) \, \mathsf{x} \, 10^{-6}$\\
			GM pulse & $100$ & $6.3(3)\, \mathsf{x} \,  10^{10} $   & $46.9(6)$ & $9.5(4) \, \mathsf{x} \, 10^{-6}$\\
			GM ramp & $70$ & $6.0(5) \, \mathsf{x} \, 10^{10} $ & $8.9(4)$ & $1.1(1) \, \mathsf{x} \, 10^{-4}$ \\
		\end{tabular}
		\end{ruledtabular}
	\end{center}
\end{table}

A comparison between the results obtained with different procedures is summarized in Table \ref{resfin}. These results are computed with a larger data set compared to each point of the previous characterizations. The constant intensity pulse sequence, represented in Fig. \ref{seq}(a), produces analogous results to the ones obtained with a red-detuned $\mathrm{D_2}$ BM sequence of $5 \; \mathrm{ms}$ duration. The \textit{PSD} after the complete GM sequence, shown in Fig. \ref{seq}(b), improves by one order of magnitude whereas in the case of BM the application of ramps to the parameters does not improve the overall performances. The optimized GM delivers samples at a temperature of $8.9(4) \; \mathrm{\mu K}$, below $4 \, T_{\mathrm{rec}}$, where $T_{\mathrm{rec}}=2.4 \; \mu \mathrm{K}$ is the recoil temperature of sodium.
We also investigated the correlation between temperature and density of the atomic sample in the range of densities between $\sim 10^{10} \; \mathrm{cm^{-3}}$ and $\sim 10^{11} \; \mathrm{cm^{-3}}$, finding no evidence of heating within our temperature sensitivity.
Our results are comparable to the current state of the art for GM cooling of other alkali atoms, where no confinement force is present. Temperatures as low as $800 \; \mathrm{nK}$, slightly above four recoil temperatures, were obtained with $^{133} \mathrm{Cs}$ \cite{Triché99}, while a \textit{PSD} of $2 \times 10^{-4}$ is reported for $^{39} \mathrm{K}$ \cite{Salomon13}. Higher \textit{PSD} can be reached combining GM with additional spatial compression stages, as reported in Ref. \cite{Elman05}, where $^{85}\mathrm{Rb}$ samples were produced with a \textit{PSD} of $1.7 \times 10^{-3}$.

\section{\label{conc}Conclusions}
We demonstrated experimentally the cooling of atomic sodium in gray molasses using the $\mathrm{D_1}$ line. We measured a nearly fourfold reduction in temperature and tenfold increase in \textit{PSD} compared with the red-detuned optical molasses sequence, and a capture efficiency higher than 2/3. The dependence of the results on the relevant parameters, as well as the significant role of coherence induced by Raman coupling between the hyperfine states, is in agreement with the results previously reported for other atomic species, confirming the solidity of the approach.
The most evident advantage of increasing the \textit{PSD} of an atomic sample is the possibility to obtain a larger number of quantum degenerate atoms after evaporation, for which sodium is particularly suited thanks to its favorable collisional properties. This will be beneficial to study the real-time dynamics of defects in BECs \cite{Freilich10,Serafini15} through partial-transfer absorption imaging \cite{Ramanathan12}.
In addition, gray molasses cooling methods should also greatly favor the realization of condensates through all-optical means.
\section{Acknowledgments}

This work has been supported by the Provincia Autonoma di Trento, by the QUIC grant of the Horizon 2020 FET program, and by Istituto Nazionale di Fisica Nucleare.



\end{document}